\documentclass{emulateapj}
\usepackage{apjfonts}
\usepackage{graphicx}
\usepackage[dvips]{color}
\usepackage{amsmath}
\usepackage{amsfonts}
\usepackage{amssymb}
\usepackage[letterpaper=true,ps2pdf=true,colorlinks=true]{hyperref}

\newcommand{\as}{$\arcsec$}
\slugcomment{Preprint, Accepted to ApJ}
\shorttitle{Optical Emission Band Morphologies of the Red Rectangle}
\shortauthors{Vijh et al.}

\begin{document}

\title{Optical Emission Band Morphologies of the Red Rectangle}

\author{Uma P. Vijh\altaffilmark{1,2}, Adolf N. Witt\altaffilmark{1}, Donald G. York\altaffilmark{3}, Vikram V. Dwarkadas\altaffilmark{3}, Bruce E. Woodgate\altaffilmark{4}, and Povilas Palunas\altaffilmark{5}}

\altaffiltext{1}{Ritter Astrophysical Research Center, The University of Toledo,Toledo, OH 43606, (awitt@dusty.astro.utoledo.edu)}
\altaffiltext{2}{Current Address: Space Telescope Science Institute, San Martin Drive, Baltimore, MD 21218, (vijh@stsci.edu)}
\altaffiltext{3}{Department of Astronomy and Astrophysics, University of Chicago, Chicago, IL 60637 (don@oddjob.uchicago.edu, vikram@oddjob.uchicago.edu)}
\altaffiltext{4}{NASA Goddard Space Flight Center, Greenbelt, MD 20771 (woodgate@uit.gsfc.nasa.gov)}
\altaffiltext{5}{McDonald Observatory, University of Texas, Austin, TX 78712 (palunas@astro.as.utexas.edu)}

\begin{abstract}
We present narrow-band images of the Red Rectangle (RR) nebula which reveal the distinct morphologies of this intriguing nebula in different optical emission bands. The morphology of the RR nebula in blue luminescence (BL) and extended red emission (ERE) are almost mutually exclusive. We also present the optical detection of the circum-binary disk of the RR in the light of the BL. The total intensities from the two optical band emissions (BL and ERE) when summed over the nebula are of comparable magnitude. Their spatial distributions with respect to the embedded illumination sources lead us to suggest that they may be attributed to different ionization stages of the same family of carriers. 
\end{abstract}

\keywords{ISM: individual(\objectname{Red Rectangle}) --- ISM: molecules --- dust, extinction --- radiation mechanisms: general --- stars: individual(\objectname{HD 44179})}

\section{Introduction}
The optical spectrum of the Red Rectangle (RR) nebula consists of several components  of comparable intensities \citep{schmidt80,wb90,vijh05}, and it is possible to use narrow-band imaging to isolate the morphological substructures where particular emissions are dominant. This is in contrast to most ordinary reflection nebulae, in which the scattered light component dominates \citep{wb90}. The special status of the RR is a consequence of the optically-thick circumstellar disk surrounding the central illuminating source and seen edge-on, which prevents forward-directed scattered radiation from reaching the observer. Thus, it is the suppression of scattered radiation by at least one order of magnitude that makes blue luminescence (BL) and extended red emission (ERE) readily observable in the RR. The latter two radiation processes are most likely the result of isotropic emission processes, which are readily observable from nebular locations where the scattered light is preferentially directed away from the observer on account of the strongly forward-throwing scattering phase function of dust at optical wavelengths.

The spectral components of the optical light of the RR are: a)~Dust-scattered radiation: to first order, this spectrum is a somewhat bluer version of that of the more luminous star in the central binary system, HD~44179, a T$_\mathrm{eff}\sim$8250~K A-giant AGB star. b)~Blue luminescence (BL) band: this emission band with a peak near 380~nm (FWHM $\sim$ 45~nm) has been attributed to fluorescence by neutral PAH molecules with 14 - 18 C~atoms. \citep{vijh04, vijh05} c)~Extended red emission (ERE): this band peaks near 670~nm (FWHM $\sim$ 180~nm)  and it is the primary cause for the unusual red color of the RR. Although observational constraints have indicated that the ERE carrier requires photons with E~$\geq$~10.5 eV for their initiation, indicating a possible association with doubly ionized molecular species \citep{witt06}, the carrier of the ERE is as yet unidentified \citep{wv04}. d)~Sharp red emission features: this is a set of relatively narrow emission bands due to  as yet unidentified molecules. They appear in a wavelength range largely coincident with that of the ERE band \citep{vanWinckel02}. It has also been  suggested that some of these bands are emission counterparts of corresponding  diffuse interstellar bands (DIBs) \citep{scarrott92}. Spectroscopic studies have demonstrated that the intensities of the emission bands b),  c), and d) are comparable to those of the underlying scattered light continuum, allowing their relative distributions to be traced. This makes it possible to constrain the physical state(s) of the respective emitters, because the  almost-edge-on geometry of the bi-polar outflow allows one to associate specific  physical conditions with regard to density, radiation field, and ionization equilibrium with  specific nebular regions. 

Our imaging study of the RR differs in significant aspects from previous efforts. Most earlier work was focused on sub-arcsecond resolution at near- and mid-infrared wavelengths with the aim to resolve the core of the RR \citep{roddier95,cruzalbes96,mekarnia98,tuthill02,miyata04}, which resulted in the well-resolved imaging of the circumstellar disk and brightness distribution of the RR in the bright innermost region. The more extended brightness distribution of the RR was studied at high resolution by \citet{cohen04}, using WFPC2 on HST. Their band-passes were limited by the filters available in WFPC2, with the F467M filter defining their shortest-wavelength band. Thus, they were unable to address the spatial distribution of the BL, which peaks below 400 nm in wavelength. They were, however, able to demonstrate the distinctly different spatial distribution of the dust-scattered radiation in the RR, which dominates the F467M band, and of the ERE, which was imaged in three different red bands (F588N, F622W, F631N). They also produced the best-resolved record of the ladder structure in the ERE crossing the outflow cones of the RR. Another important mapping effort is that of \citet{breg93}, who imaged the extended brightness distribution of the RR in the two aromatic emission features at 3.3 \micron\ and 11.3 \micron. They demonstrated that the 11.3 \micron\ brightness distribution followed the extended bipolar structure of the RR, while the 3.3 \micron\ isophotes were more concentrated near the center of the nebula. Our goal was to pursue the large-scale brightness distribution in the RR of specific optical band emissions, which we attempted to isolate with narrow-band filters and still narrower Fabry-Perot bandpasses, while the angular resolution was limited by seeing ($\sim 1$ \as). Our work includes, therefore, a first effort to image the BL in the RR.

\section{Observations and Reductions}
The data were obtained at the Apache Point Observatory on January 7 -- 11, 2005, using the Goddard Fabry Perot (FP) on the 3.5-m telescope. Images of the RR were obtained using narrow-band filters centered at 3934~\AA\ ( FWHM $\sim$ 27~\AA), 4050~\AA\ (FWHM $\sim$ 57~\AA), 5700~\AA\ (FWHM $\sim$ 135~\AA), 6400~\AA\ (FWHM $\sim$ 135~\AA).  The nights were partially cloudy and only relative fluxes could be obtained from these images. We used the FP etalon to image the narrow emission features (item d of the list in the introduction), but could not get good enough exposures due to clouds. This paper deals only with the pre-pass filter images taken during the run.

Data reduction was carried out using standard IRAF tasks. All images were trimmed to a size of 561~$\times$~561 pixels or 3$\farcm$4 $\times$ 3$\farcm$4. The darks, twilight flats and object frames were corrected using an average bias frame. The flats and object frames were then dark-corrected using a combined dark frame. The flat frames for each filter were obtained by moving 30\as\ between successive exposures to avoid coherent addition of bright stars. These flat frames were then median-combined for each filter. Dark corrected object frames were then flat-fielded using the combined twilight flat for each filter. Object exposures were taken in sets of five, in a pentagram pattern of pointings. Object frames with the best seeing were selected and median-combined using tasks \emph{imalign} and \emph{imcombine}. Table~\ref{obs-table} shows the exposure times for each filter. Stars in the frames were used to register and shift the images: this worked for all filters except the 3934~\AA\ filter. As no stars other than the central extended source were visible on these frames, the center of the extended source was used to shift and co-add the images. This was found to be reliable after the fact as one of the next-brightest stars in the field now appeared in the co-added image, albeit with extremely low counts. All images were further cropped to 200~$\times$~200 pixel or 73\as~$\times$~73\as\ to best show the RR. In the absence of calibration, images were scaled using background counts at large offsets from the central bright source before computing the ratio images discussed in the next section.

\begin{table}
\caption{Narrow-band observations of the RR nebula.}
\label{obs-table}
\centering
\begin{tabular}{cccl}
\hline \hline \\
Filter & Filter width (\AA) & Exposure (s)& Dominant Component\\
\hline  \hline \\
3934 & 27 & 5$\times$900 & BL band \\
4050 & 57 & 5$\times$120, 1$\times$180 & Scattered \\
5700 & 135 & 5$\times$ 120 & Scattered + ERE \\
6400 & 135 & 4$\times$60, 1$\times$180 & ERE + Scattered \\
\hline \\
\end{tabular}
\end{table}

\section{Results}
\subsection{Blue Luminescence in the Red Rectangle}
\begin{figure}
\centering
\includegraphics[width=3.3in]{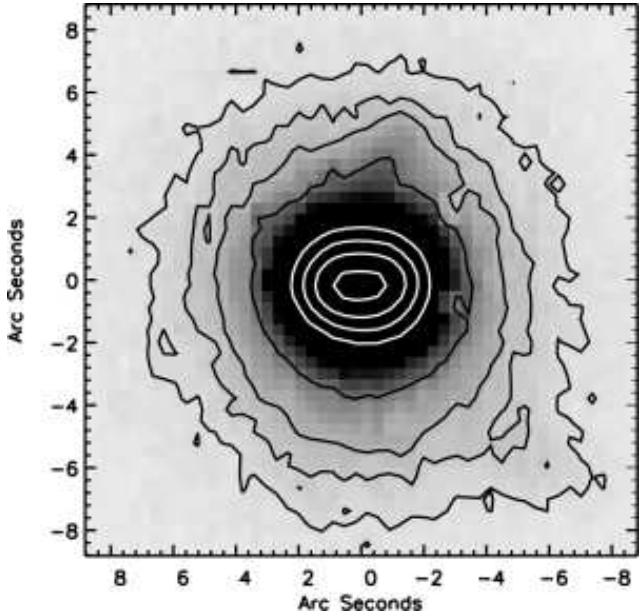}\\
\caption{Image at 3934~\AA\ with overlaid contours (at 0.5\%, 0.75\%, 1\%, 2\%, 10\%, 20\%, 40\%, 80\% of maximum intensity) showing elongated profiles along the disk tracing the BL distribution.}
\label{3934cont}
\end{figure}

\begin{figure}
\centering
\includegraphics[width=3.3in]{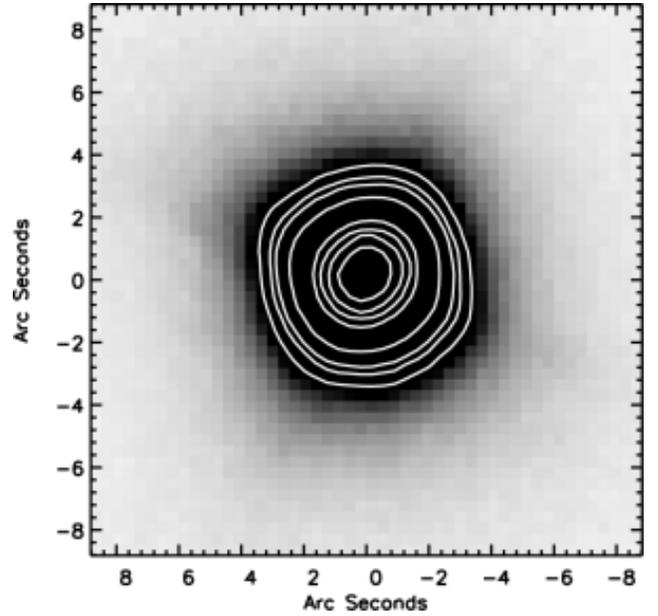}
\caption{Image at 4050~\AA\ with overlaid contours (at 0.5\%, 0.75\%, 1\%, 2\%, 10\%, 20\%, 40\%, 80\% of maximum intensity) showing circular scattered light profiles.}
\label{4050cont}
\end{figure}

\begin{figure*}
\centering
\includegraphics[width=3.3in]{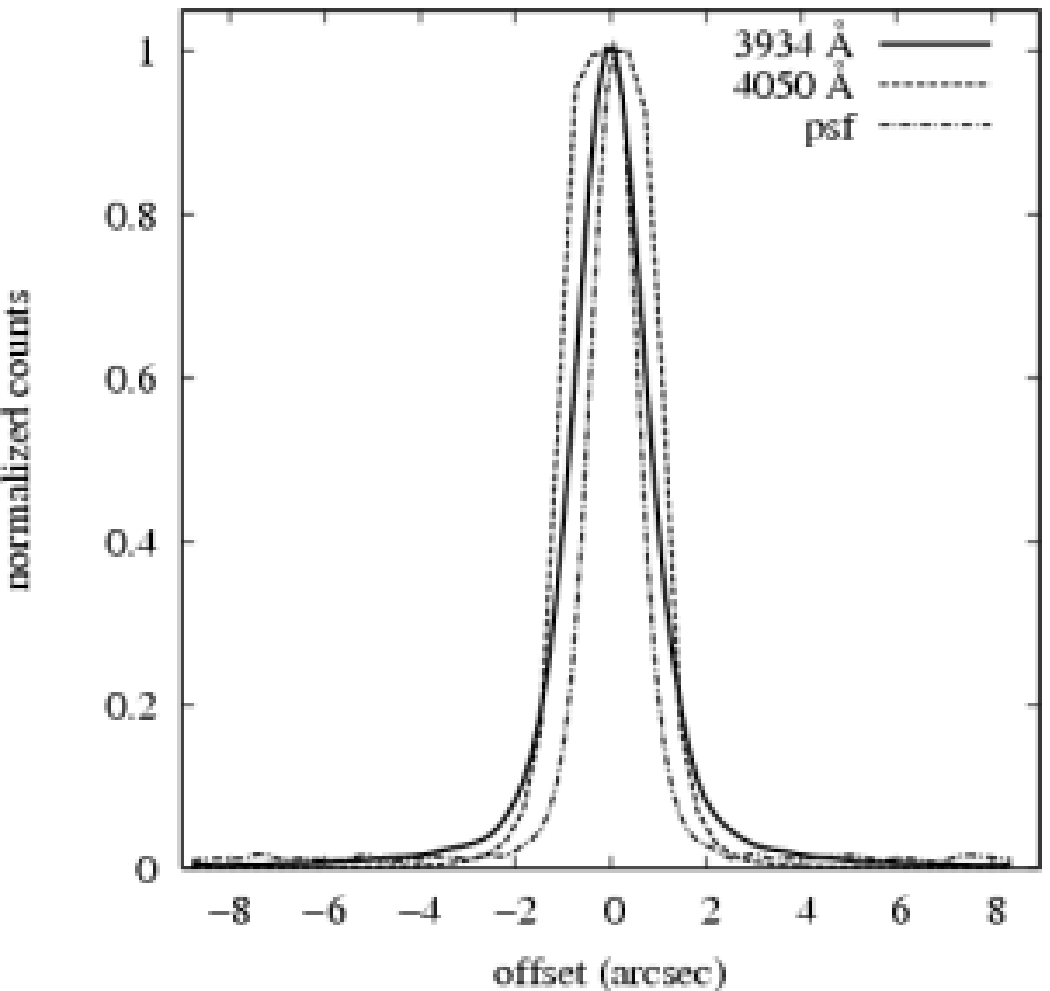}\includegraphics[width=3.3in]{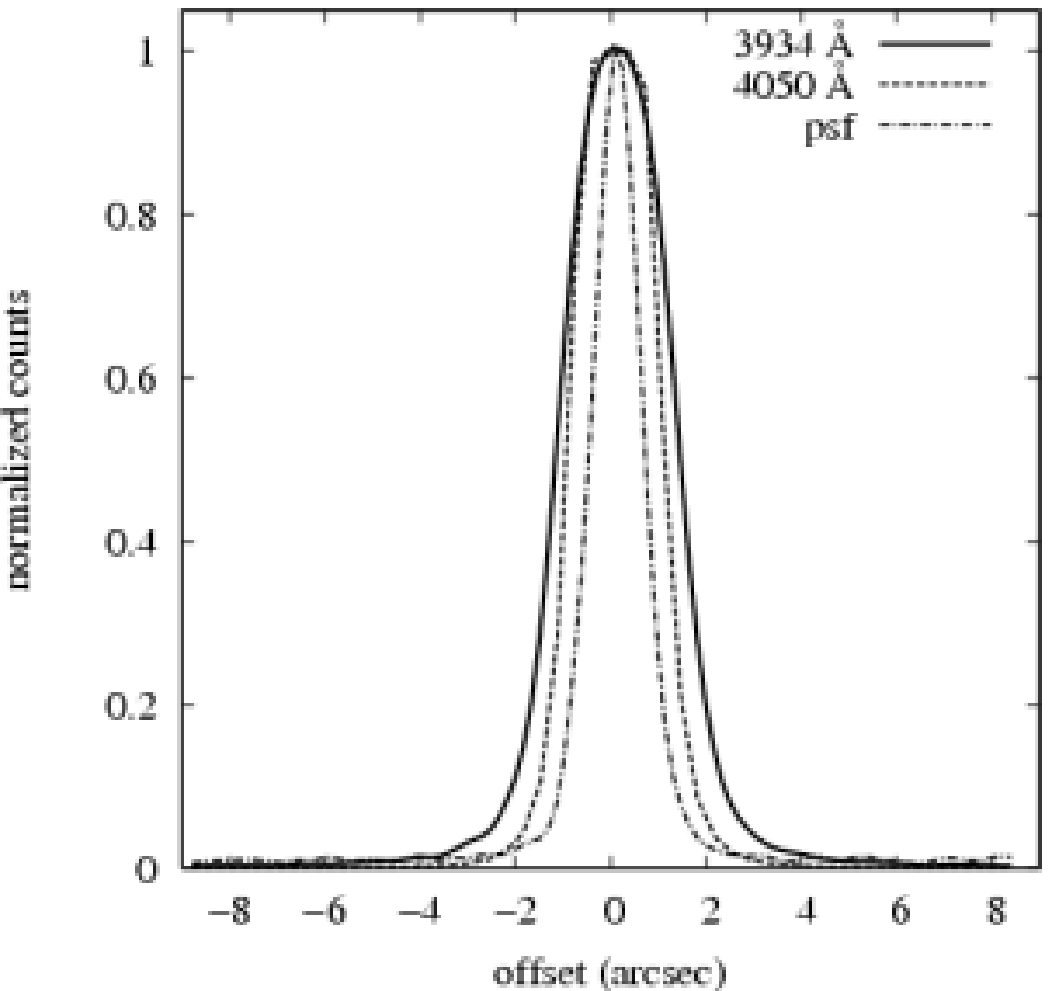}
\caption{Intensity traces with the point-spread function (psf) of the detector. N-S (left) and E-W (right) traces of the intensity distributions at 3934~\AA\ and at 4050~\AA\ are compared with the psf of the detector, derived from stellar images in the field imaged in the 4050~\AA\ band. In the N-S direction the inner core of the 3934~\AA\ image, which is more strongly affected by the BL than the 4050~\AA image, is barely resolved, in contrast to the broader 4050~\AA\ image which hints at a comparatively broader distribution. In the E-W direction, both BL-dominated (3934~\AA) and scattered light-dominated (4050~\AA) intensity profiles of the RR are well resolved, and the ratio of BL to scattered light is increasing with offset from the center in both E and W directions.}
\label{psf}
\end{figure*}

The two narrow-band images at 3934~\AA\ and 4050~\AA\ (Figure~\ref{3934cont} and \ref{4050cont}) show the appearance of the RR in  the wavelength region of the BL band. The former is much closer to the peak of the BL  emission, while the latter covers the long-wavelength tail of the BL band. This causes the $\mathrm{\lambda}$4050 image to include a larger contribution, in relative terms, from the scattered light component of the spectrum. Figure~\ref{3934cont} is an image of the RR taken with the 3934 filter, superimposed with intensity contours. The most remarkable feature of this image is the fact that the inner isophotes are consistently elongated in the E-W direction. A check on stellar images in the  3934~\AA\ frames confirmed that this elongation is real and not the result  of guiding errors  in the individual 900 s exposures. This E-W elongation in the blue image was confirmed in a subsequent observation obtained at APO on December 16, 2005. As these observations too, are not calibrated and less deep than those discussed in this paper, they have not been used in the analysis. The E-W elongation is further illustrated in Figure~\ref{psf}, where we compare N-S and E-W traces of the intensity distributions at 3934~\AA\ and at 4050~\AA\ with the point-spread function (psf) of the detector, derived from stellar images in the field imaged in the 4050~\AA\ band. In the N-S direction the inner core of the 3934~\AA\ image, which is more strongly affected by the BL than the 4050~\AA\ image, is barely resolved, in contrast to the broader 4050~\AA\ image. This demonstrates that the BL is exhibiting a strong intensity peak at the position of the circumstellar disk, with a steep intensity gradient toward the regions above and below the disk. The 4050~\AA\ N-S profile, on the other hand, exhibits a comparatively broader distribution in this direction. A quite different result emerges from the traces in E-W direction. Both BL-dominated (3934~\AA) and scattered light-dominated (4050~\AA) intensity profiles of the RR are well resolved, and the ratio of BL to scattered light is increasing with offset from the center in both E and W directions. This set of traces confirms this same trend observed in the RR with long-slit spectroscopy with E-W oriented slits by \citet{vijh05}, which  showed that the BL emission exhibits a strong preference for the parts of the RR that are located in the disk, shielded from direct irradiation by the star. 

The E-W elongation of the BL disk is consistent with the identification of the BL  as originating from fluorescence by small (3- to 4-ring) \emph{neutral} PAH molecules as proposed by \citet{vijh05}. Such molecules are photo-excited by stellar photons of $\sim$~4~eV energy, which can diffuse into the outer disk regions via scattering. The same molecules would be photo-ionized by 7.5 eV photons, which are present in regions of the RR with direct lines-of-sight to the  central sources, e.g. the outflow cones. Once ionized, the same molecules assume a different electronic energy level structure and are no longer able to fluoresce in the blue spectral range. \citet{vijh05} examined other potential carriers of the BL band such as silicon nanoparticles, hydrogenated amorphous carbon and silicon carbide, and found their spectra to be inconsistent with the obsereved BL spectral characteristics. We believe, therefore, that the 3934~\AA\ image is showing us  emission from the outer portions of the optically thick disk surrounding the central stars  in the RR, i.e regions where the local radiation field causes the ionization equilibrium to  be preferentially neutral. We  note that the isophotes of the 4050~\AA\ image, shown in Figure~\ref{4050cont} are distinctly circular in the inner  region of the image, in contrast to the elongated 3934~\AA\ isophotes. This reflects the fact that dust-scattered light is dominant here. The outer isophotes of the 4050~\AA\ image  become almost square, with the two sides being parallel to the polar axis of the RR nebula. While  the BL image at 3934~\AA\ shows slight intensity enhancements at the base of the outflow  walls on the W-side of the nebula, no such enhancements are apparent in the 4050~\AA\ image, despite the fact that this is a deeper exposure. The dimensions of the RR disk implied by the narrow-band imaging observations are in agreement with the results of \citet{buj03,bujarrabal05} who  mapped this disk in CO emission with the IRAM interferometer array. Their map shows the disk having an angular extent of $\sim$~6\as. The 3934~\AA\ data show elongated isophotes over an  extent of 6.5\as, which is in excellent agreement, taking into account the $\sim$~1\as\ seeing disk of the optical data. 

A perplexing aspect of the 3934~\AA\ image is the orientation of the disk (PA~$\sim$~90$^\circ$) compared to the orientation of the polar axis of the RR (PA~$\sim$~11$^\circ$). If the disk were  exactly perpendicular to the polar axis, a PA of 101$^\circ$ would have been expected. This PA of 101$^\circ$ appears to be in agreement with disk images obtained in absorption at near-IR \citep{tuthill02}. However, these near-IR images show the disk only over an extent of 200 mas to 500 mas, while the elongated emission recorded in the 3934~\AA\ image extends over a much larger angular extent, larger by a factor of 10 to 20. The CO observations \citep{bujarrabal05} also indicate that the CO disk is perpendicular to the RR axis and is aligned with the disk orientation derived from the near-IR and optical images. To establish that we would indeed be able to see a 11$^\circ$ rotation in our image we made a high-res test bar (0.3\as\ high, 4\as\ long, PA 101$^\circ$) to simulate the RR disk. We degraded this test-image to the observed resolution by using the observed psf (at 4050~\AA) and the scattered light profile of the RR from the 4050~\AA\ image. The contours of the simulated bar clearly run at PA 101$^\circ$ and not in the pixel direction. It appears likely, therefore, that we are recording emission from the outer reaches of the circumstellar disk, with the intensity distribution strongly influenced by the excitation conditions for the molecules responsible for the BL.

\subsection{Extended Red Emission}
\begin{figure}
\centering
\includegraphics[width=3.3in]{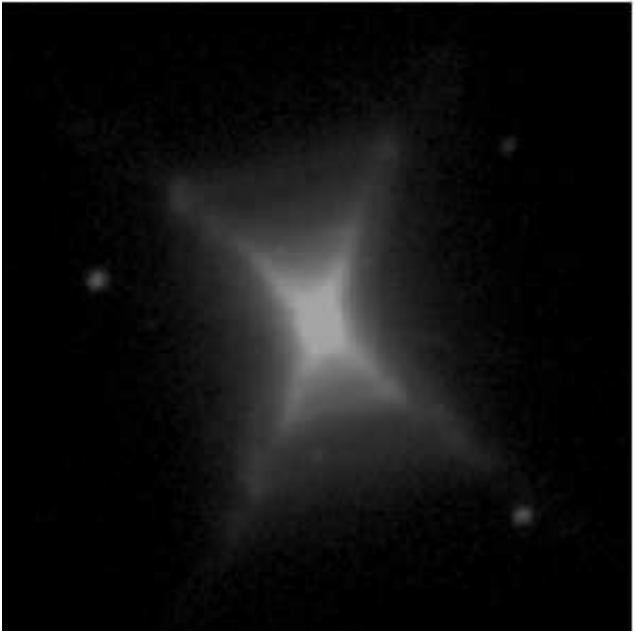}
\caption{RR image at 6400~\AA, with 135~\AA\ wide filter. The image at this wavelength is dominated by ERE. North is to the top and east to the left. Image is 73\as~$\times$~73\as. Note the incipient appearance of the ladder structure that is so clear in the HST images, and its sharpening, along with the bright knots at the ends of the ladder steps,  in Fig.~\ref{64div40} (right image), when the scattered light is divided out.}
\label{fig-ere}
\end{figure}

\begin{figure*}
\centering
\includegraphics[width=3.3in]{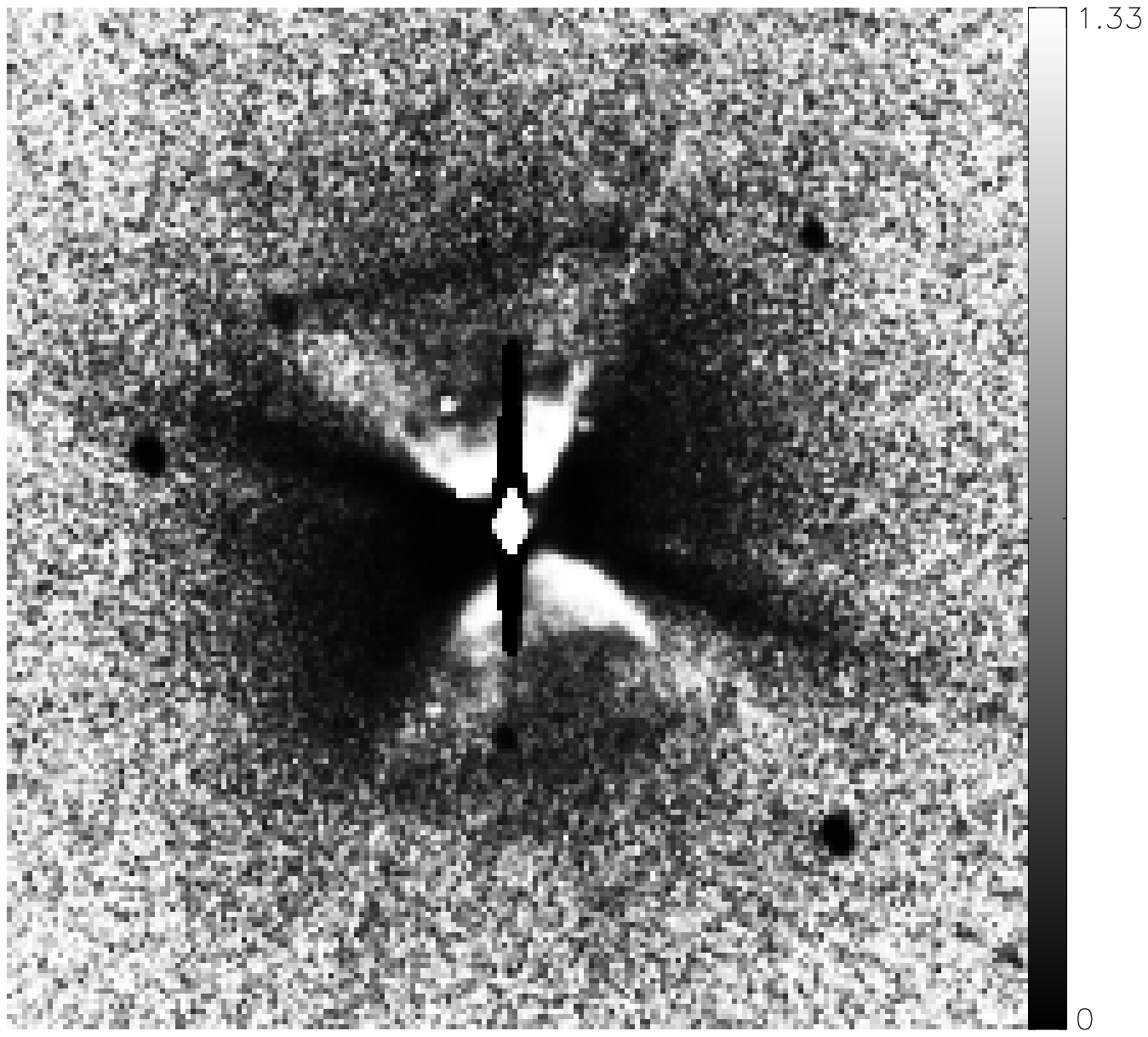}\includegraphics[width=3.3in]{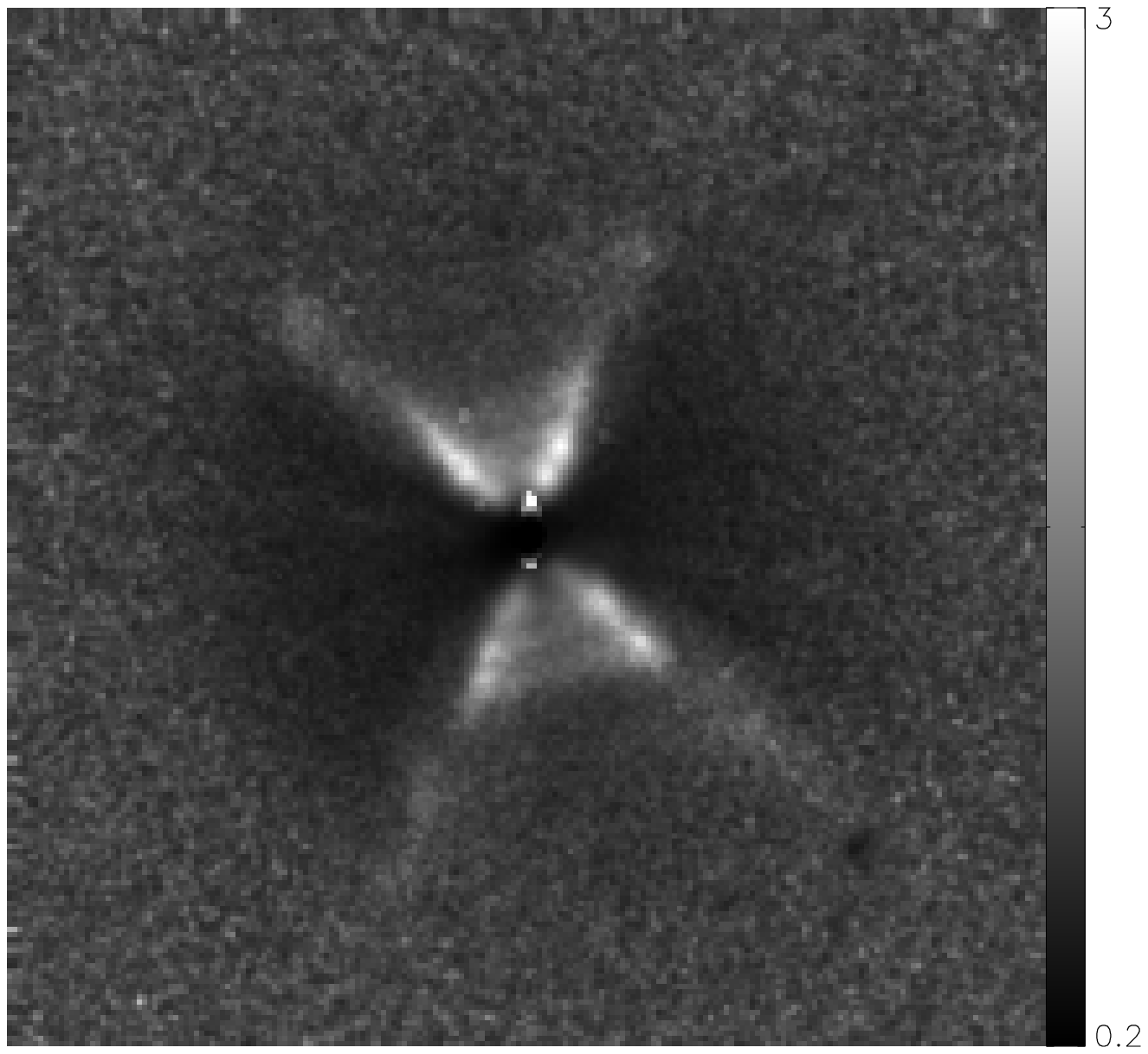}
\caption{Two ratio images. (left panel): 6400~\AA\ image divided by 5700~\AA\ image. The dark, vertical band in the center is caused by bleeding of the over-exposed stellar image. (right panel): 6400~\AA\ image divided by 4050~\AA\ image. Both images are 73\as~$\times$~73\as, and north is to the top and east to the left. The 5700~\AA\ and 4050~\AA\ bands are dominated by scattered light, which clearly dominates outside the outflow cones, whereas the ERE dominates in the cavity walls.}
\label{64div40}
\end{figure*}

The 6400~\AA\ image is dominated by the ERE (see Fig.~\ref{fig-ere}). When compared  to images at wavelengths $<$ 5000~\AA, this image demonstrates that the X-shaped  structures (variably referred to as ``arms'' or ``whiskers'' in the RR literature) are a result of the ERE band. We interpret these structures as the walls of the internally illuminated  bi-polar outflow cavities, seen in projection. When divided by the 5700~\AA\ ``continuum'' image, the resulting 6400/5700 ratio image, Figure~\ref{64div40} (left) convincingly demonstrates that the ERE is preferentially emitted by material in and within the walls of the outflow cavities, and is extremely weak outside these regions. This is demonstrated even more clearly in the ratio image 6400/4050, Figure~\ref{64div40} (right). The 5700~\AA\ image is  slightly overexposed, leading to bleeding from the stellar image and to the appearance  of diffraction spikes that are not visible on the other images. Figure~\ref{64div40} is basically a  display of the ratio of ERE to dust-scattered radiation in the RR. We can interpret the relative confinement of the ERE to the outflow cavities and their walls in one of two ways: either the carrier of the ERE exists only in the outflow cavities and its walls, or the excitation requirements for the ERE restrict this emission process to regions which can receive the photons required for ERE excitation from the central sources. Either interpretation is consistent with the findings of \citet{cohen04}, who demonstrated that the characteristic X-shaped structure of the RR is visible in several red bands dominated by the ERE but is absent in the F467M image.

\section{Discussion}
\subsection{Excitation of the ERE}
The two possible interpretations of the apparent confinement of the ERE to the walls of the outflow cones of the RR can actually be combined in the light of the results of a recently completed investigation \citep{witt06} of the ERE in the reflection  nebula NGC~7023.  This study concluded that the ERE results from a two-step process. Step 1 consists of the creation of the ERE carrier. This involves the ionization (double ionization in case of PAHs) of the ERE carrier progenitor by photons with energies 10.5~eV $<$ E $<$ 13.6~eV. This step can proceed only inside the outflow cavity and its walls with a direct line of sight to the central stars, in particular to the source of far-ultraviolet photons associated with them. Step 2 consists of the pumping of the ERE carrier, followed by fluorescence. The pumping relies on the presence of abundant optical/near-UV photons from the luminous AGB star HD 44179 located at the center of the RR. 

The evidence for a source of ionizing far-ultraviolet radiation near the center of the RR, although indirect, is strong. The central source in the RR has been identified as a binary \citep{waelkens96, bakker98,menshchikov02} with a orbital perios of 320 days. While hidden from direct view at optical wavelengths, the center of the RR has been mapped at cm-radio wavelengths by \citet{jura97} and was found to exhibit a spectrum consistent with a compact HII region. This is further supported by the observations of scattered emission in the HeI recombination line at 1.0830 $\mu$m by \citet{kelly95}, of strong hydrogen Balmer alpha line emission \citep{jura97}, and a host of other permitted and forbidden atomic emission lines \citep{hobbs04}, all associated with the central region of the RR. 

The source of this ionization has remained a subject of some controversy. The absence of an observed far-UV excess in the spectrum of HD~44179 allowed \citet{menshchikov02} to place an upper limit of 100 L$_\odot$ on the luminosity of the UV source, which, together with information on the mass of the secondary star in HD 44179, led them to suggest a hot helium white dwarf as the source in question. While it seems clear that the photosphere of the A1 III primary in HD 44179 can be ruled out as the source of the ionizing radiation, \citet{jura97} discussed the possibility that accretion on either the primary or the secondary star in HD 44179 or onto an accretion disk surrounding the secondary could be the source of the ionizing radiation. Regardless of whether due to a hot white dwarf or accretion, UV continuum radiation in the energy range 10.5~eV $<$ E $<$ 13.6~eV will escape from the central region preferentially in the direction perpendicular to the circumstellar disk, where it can be absorbed by material in the walls of the outflow cones to lead to the initiation of the ERE carrier.

While optical/near-UV photons from the A1 III primary of HD~44179 permeate other parts of the nebula as well, as demonstrated by the RR images at wavelengths dominated by scattered light, they can excite the ERE only in those places where  the earlier initiation has created the carriers. Neutral PAHs, such as those postulated in the RR disk shadow will also absorb these optical/near-UV photons and respond with blue fluorescence. If PAHs are responsible for the ERE, the initiation constraint found in  NGC~7023 suggests that PAHs are ionized to the di-cation stage before they luminesce in the ERE band. Quantum-chemical calculations of the electronic energy level structure  of PAH di-cations show the presence of strong absorption bands shortward of $\sim$ 6000~\AA\ \citep{witt06}. Whether or not PAH di-cations fluoresce in the ERE band ($>$ 6000~\AA) is currently the subject of laboratory investigations.     

\subsection{Chemical Environments in the Red Rectangle}
\citet{waters98} used ISO-SWS spectra from a large aperture which included most of the RR nebula (14\as~$\times$~20\as) for their study of the spectrum of the RR in the mid-IR. These spectra showed crystalline silicate emission features and CO$_2$ absorption features.  They also compared broad-band 10~\micron\ (N-band) and narrow-band 11.3~\micron\ images, the latter emission suggested to arise from a mixture of neutral and ionized PAHs \citep{allamandola99,bakes01}. These images showed that while the 10~\micron\ flux distribution was centrally peaked, the 11.3~\micron\ image showed an extended distribution. \citet{breg93} noted that the 11.3~\micron\ image showed a narrow-waisted, bipolar shape similar to the ERE image. In contrast, their 3.3~\micron\ (arising from predominantly neutral PAHs) image was more centrally peaked and lacking in extended emission in the outflow regions, although the 3.3~\micron\ is by no means confined to the disk region only.

Based on the N-band and 11.3~\micron\ images \citet{waters98} claimed that the disk in the Red Rectangle is made up of oxygen-rich material produced in an earlier mass-ejection epoch, while the bi-polar outflow lobes are carbon-rich matter produced during the more recent outflow. However, the spatial resolution of their spectra was insufficient to discriminate between the emission from the disk and outflow regions. Their claim relies mainly on the argument that the present-day outflow is C-rich and that crystalline silicates are unlikely to form in an outflow. They concluded, therefore, that the O-rich material must be in a stable configuration like the disk. Our observations raise questions about such a simple division into two chemically different environments. If the BL arises from PAHs \citep{vijh04,vijh05} our observations demonstrate that such carbon-based molecules are abundantly present in the disk as well. \citet{mulas06} compute the absolute fluxes for the vibrational IR emission from small, neutral PAHs carriers proposed by \citet{vijh05} and show that if anthracene were responsible for the BL, then it would almost completely account for the observed flux in the 3.3 \micron\ band. Taking into consideration that only 10-20\% of the Si needs to be crystalline to account for the spectra \citep{waters98,kemper04} overcomes the argument for all the C-rich material to be confined to the outflow cones and the disk be completely formed by O-rich material. Our data are not inconsistent with the picture of two successive mass ejection epochs, a C-rich one following an earlier O-rich one, proposed by \citet{waters98}. They do suggest, however, that small PAH molecules (if they are the BL carriers) formed during the present C-rich ejection have also entered the space outside the outflow cones. It is in this part of space, especially in the shadow of the circum-binary disk of the RR, where such molecules are protected against ionization and are thus able to fluoresce in the blue part of the spectrum.

In summary, the narrow-band morphology of the RR in the BL, ERE, near- and mid-IR emission bands gives us further insight into the distribution of chemical environments in this system. The disk regions show spectral emission features likely produced by neutral PAHs (3.3 \micron\ emission, BL) in addition to crystalline olivine features seen at 11.27~\micron\ \citep{miyata04}. The outflow regions show emission from 11.3~\micron\ \citep{breg93, miyata04} and from 0.6~\micron\ (the ERE). The former may probably originate from neutral and/or singly ionized molecules and the latter from doubly ionized PAH species.

\section{Conclusions}
\begin{enumerate}
\item The morphology of the BL and ERE emissions in the RR nebula are almost mutually exclusive. The ERE is strongly enhanced along the walls and in the interior of the bipolar outflow cones of the RR nebula, while the BL is strongest in those regions that are located in the shadow of the circumstellar disk surrounding the central source, HD 44174.

\item The presence of a circumstellar disk, viewed edge-on, divides the space around HD 44179 into two environments characterized by radiation fields of greatly different intensity and hardness. This leads us to suggest that the ERE and the BL arise from different ionization stages of the same family of carriers, such as PAH molecules. In case of PAH molecules, a large range of sizes and species are probably present. The BL preferentially detects small, neutral molecules while the mid-IR aromatic emission features arise from both neutral and ionised species. The ERE, if due to PAHs, arises from doubly ionized molecules with masses $>$ 200 amu.

\item  Our observations at 3934~\AA\, near the peak of the BL band, show an intensity distribution elongated in E-W direction, with a maximum at the position of the circumstellar disk of HD 44179. If this structure is associated with the outer portions of the disk, its alignment differs from that of the inner disk observed in absorption at near-IR wavelengths and the disk seen in CO emission by 11 degrees.

\item Our observations add to the growing evidence for greater chemical complexity of the RR nebula than previously proposed.
\end{enumerate} 

\acknowledgements
This research has been supported by the NSF Grant AST 0307307 to The University of Toledo. VVD's research is supported by NSF Award 0319261. Data presented here are based on observations obtained with the Apache Point Observatory 3.5-meter telescope, which is owned and operated by the Astrophysical Research Consortium. We would also like to thank Steve Howell for facilitating the loan of the KPNO filter used for the BL observations. This research has made use of NASA's Astrophysics Data System Bibliographic Services, the SIMBAD database, operated at CDS, Strasbourg, France and also of the Aladin image server.

\bibliography{thesis_ref}

\end{document}